\documentclass[a4paper]{article}

\usepackage{INTERSPEECH2021}
\usepackage{todonotes}
\usepackage{cleveref}
\usepackage{graphicx}
\usepackage{subfig}
\usepackage{cite}
\def\accstatic/{56.96} % on 10% test data as per CSV from Franzi, 10k
\def\accneural/{97.10}  % on 10 test data as per CSV from Franzi, 10k

\usepackage{color, colortbl}
\definecolor{LightCyan}{rgb}{0.88,1,1}
\usepackage{float}

\title{Attacker Attribution of Audio Deepfakes}
\name{Nicolas M. M\"uller$^{1*}$\thanks{$^*$Equal contribution}, Franziska Dieckmann$^{2*}$, Jennifer Williams$^3$}
\address{
  $^1$Fraunhofer AISEC, Germany\\
  $^2$Technical University of Munich, Germany \\
  $^3$University of Southampton, UK
  }
\email{nicolas.mueller@aisec.fraunhofer.de, franziska.dieckmann@tum.de, j.williams@soton.ac.uk}

\begin{document}

\maketitle
\begin{abstract}
% \todo{JW has fixed length to be under 200 words. TODO update abstract in submission system}
   Deepfakes are synthetically generated media often devised with malicious intent. They have become increasingly more convincing with large training datasets advanced neural networks. These fakes are readily being misused for slander, misinformation and fraud.     For this reason, intensive research for developing countermeasures is also expanding. However, recent work is almost exclusively limited to deepfake detection - predicting if audio is real or fake. This is despite the fact that attribution (\textit{who created which fake?}) is an essential building block of a larger defense strategy, as practiced in the field of cybersecurity for a long time. This paper considers the problem of deepfake attacker attribution in the domain of audio. We present several methods for creating \textit{attacker signatures} using low-level acoustic descriptors and machine learning embeddings. We show that speech signal features are inadequate for characterizing attacker signatures. However, we also demonstrate that embeddings from a recurrent neural network can successfully characterize attacks from both known and unknown attackers. 
   Our attack signature embeddings result in distinct clusters, both for seen and unseen audio deepfakes.
   We show that these embeddings can be used in downstream-tasks to high-effect, scoring \accneural/\% accuracy in attacker-id classification.
%   Our attack signature embeddings result in distinct clusters and we present our technique for how to correctly attribute new, unseen audio deepfakes from previously unknown attackers. 
%   We present results from the ASVSpoof 2019 and 2021 dataset, achieving \accneural/ percent accuracy in a supervised downstream classification task. % on the 2019 dataset.
\end{abstract}
\section{Introduction}
Deepfakes are synthetically generated media content often created and delivered with malicious intent. Very large datasets, increasing compute power, and the use of advanced neural networks have made it possible to create very convincing deepfakes. Deepfakes are already being misused in everyday life for slander~\cite{spivak2018deepfakes}, fake news~\cite{botha2020fake} and even financial fraud~\cite{AVoiceDe77:online}. For this reason, the detection of deepfakes (or spoofs) is a major subject of current research, both in the domain of video and audio.
Especially in the domain of audio, most research is currently focused only on the \textit{detection} of fake voice recordings. For example, the popular ASVspoof Challenge~\cite{asvspoof2019challenge, asvspoof2021challenge}, only considers detection tasks for audio deepfakes but does not handle attacker \textit{attribution}. Attribution is more challenging than detection since the latter is essentially a binary classification problem (i.e., speech is either real or fake), but attribution requires a much finer gradation than `real or fake'. In this paper we present work on \textit{attacker attribution} using a technique that allows us to develop special signatures that represent particular attackers.

Working on attacker attribution is important because it allows a better understanding of the threat landscape. In cybersecurity, for example, attribution of attacks is already an important part of threat mitigation strategy~\cite{rid2015attributing, egloff2019public, maglaras2019threats, skopik2020under, egloff2021publicly}.
Our paper deals with the attribution of audio deepfakes through the creation of attacker signatures and we make three main contributions. 
    % \item Study problem of audio deepfake attribution.
(1) We introduce and evaluate a signature based on 16 low-level acoustic features such as pitch, jitter, etc. with respect to how well the signature performs in characterising audio deepfake attackers.We show that these simpler features have only limited applicability.
(2) We show that neural embedding signatures can successfully cluster deepfake speech audio with respect to attacker ID from a labeled corpus. We evaluate the neural embeddings on a large audio deepfake corpus and show that the resulting embeddings allow simple downstream models to achieve high classification accuracy.
(3) We provide insight into how our experiments could be used for developing cybersecurity tools. 

\section{Related Work}
%\subsection{Deepfakes}

The term deepfake originated in 2017 from a Reddit user who posted face-swapped pornographic videos \cite{reddituser:online}.
Since then, media content forgery using machine learning has developed rapidly%~\cite{westerlund2019emergence, korshunova2017fast}, 
which in turn motivated a flurry of research on deepfake detection~\cite{wen2015face, amerini2019deepfake} and corresponding challenges such as Facebook's Deepfake Detection Challenge~\cite{dolhansky2019deepfake}. In the domain of audio and voice, which is the focus of this paper, these fakes are commonly referred to as `spoofs'~\cite{asvspoof2019challenge, asvspoof2021challenge}.

%\subsection{Audio Synthesis / Deepfake Creation}
Audio spoofs can be created in several ways, and one such way involves using \emph{text-to-speech} (TTS) synthesis models such as Tacotron~\cite{tacotron1} combined with a vocoder such as Griffin-Lim~\cite{griffin1984signal} or \emph{neural vocoders} such as WaveNet~\cite{oord2016wavenet}. %These models take as input a dataset of transcribed speech alongside speech audio and learns a mapping between graphemes from the transcription and regions of a speech spectrogram. The speech audio must be converted to time-frequency spectrograms via a short-time Fourier transformation (STFT).
%At inference time, Tacotron recursively synthesizes a spectrogram given a target sentence. The spectrograms can be synthesized into speech using either traditional algorithms such as Griffin-Lim~\cite{griffin1984signal} or \emph{neural vocoders} such as WaveNet~\cite{oord2016wavenet}. 
With either of these vocoder methods, the spectrogram is inverted and a raw waveform of speech can be obtained.
Tacotron and its successor Tacotron 2~\cite{tacotron2} have triggered considerable follow-up research, which optimizes either synthesis quality~\cite{choi2020attentron}, minimizes inference time~\cite{ren2019fastspeech} or allows for controllable prosody~\cite{wang2018style}.
However, TTS synthesis technology can also be used to create deepfakes of a voice by cloning a person's voice and controlling what they say without consent. 

%controlling the content of what they say, and without requiring a significantly large number of voice samples. 
%Another type of TTS synthesis, called \emph{multi-speaker TTS}, requires only a small amount of a target speaker voice (as short as a few seconds) to condition a modified Tacotron architecture~\cite{jia2018transfer}. 
%This can effectively result in cloning a person's voice and controlling what they say without their consent.

%\subsection{ASVspoof and Audio Deepfake Detection}
The threat that deepfakes pose to society has triggered research in the field of audio spoof detection and led to the establishment of a large biennial challenge called the ``Automatic Speaker Verification and Spoofing Countermeasures'' (ASVspoof) series~\cite{asvspoof2019challenge, asvspoof2021challenge}.
In this challenge, the organizers provide large datasets which contain both authentic speech (i.e. spoken by a real human) and spoofed speech (i.e. fake speech from TTS synthesis, voice conversion, or neural vocoder systems).
To take the progress of TTS synthesis and voice conversion into account, new benchmark datasets are released every other year (c.f.~\ref{ss:data}).
This challenge has incited many papers which have advanced state-of-the-art in audio spoof detection~\cite{chintha2020recurrent, alzantot2019deep, chettri2019ensemble, wang2020densely}.
However, the community has perhaps been too optimistic about the detection rate of new audio spoofs not included in the training dataset, as it has been shown that some datasets contain artifacts that make classification tasks trivial~\cite{muller2021speech} due to the amount of silence padding in a waveform. 

%\subsection{Attacker Attribution}
Recording environment signatures were first created for modeling physical attack characteristics in the ASVspoof 2019 Challenge \cite{williams2019speech}. While the environment embeddings did help with attack detection, they did not identify particular types of attacks. A related problem is attacker attribution, i.e. answering the question: 'Who created this audio spoof?'.
Attribution requires the creation of attacker signatures which remain constant over all of the attacker's audio spoofs.
In this case, an attacker is identified by the specific speech synthesis setup they use which includes not only the choice of architecture but training data and hyperparameters as well.
In the domain of video deepfakes, there is very little prior work on attribution~\cite{khoo2021deepfake, zhang2020attribution}.
In the domain of audio deepfakes the work of \cite{zhao2018spoofing} uses clustering~\cite{zhao2018spoofing} for spoof \emph{detection}, but there is no prior work on attacker attribution.

\section{Attacker Signatures}\label{s:methodology}
Recall that learning attacker signatures is a different technical problem than detecting deepfakes. The primary difference involves the assumption that the audio is already known to be a deepfake and therefore the task is to attribute each deepfake to the system of origin (vocoder, TTS system, voice conversion system, etc). 
This section describes two main approaches for creating attacker signatures: extracting low-level signal features and learning embeddings from a neural network.

We define the attacker signatures as follows:
For attackers $A_0, A_1, ...$ and raw audio waveforms $W_0, W_1, ...$, we are interested in a a signature $f$ s.t. $f(W_l) = f(W_k)$ if and only if $W_l, W_k$ originate from the same attacker.
We can relax the problem and write $f(W_l) \approx f(W_k)$. 
This requires a notion of distance in $f(W)$-space, for example $L_2$ or cosine similarity.
Note that we use the notation of $A_0$ to indicate bonafide speech (i.e. authentic, non-spoofed audio waveforms) since this notation follows a similar pattern as the labeled ASVSpoof dataset that we used in our experiments.

\subsection{Low-Level Features}\label{ss:conventional_feats}
A natural first approach for creating attacker signatures is using features that can be computed directly from raw audio.
Related work has shown that even simple features, such as the length of audio silence, can suffice to verify authenticity~\cite{muller2021speech}.
We extract the following low-level signal features from raw audio using \emph{Parselmouth} ~\cite{parselmouth} with \emph{Praat}~\cite{Praatdoi65:online} and \emph{Pydub}~\cite{robert2018pydub}. The extracted features are combined into 16-dimensional vectors and used in our analysis for clustering and classifying attackers.

\textbf{Fundamental frequency.} The mean, minimum and maximum F0 (as extracted by Parselmouth), including the standard deviation and mean absolute slope (MAS). The pitch MAS is the frequency of vibration of the sound waves. All pitch values were calculated from the utterance-level waveform. This feature was chosen because it is well-known that machine-generated speech often lacks naturalness for prosody with variation among different types of systems \cite{gao2021detection}. 

\textbf{Distortion and shimmer.} We used jitter (i.e. time distortions from the digital audio signal), and shimmer, where shimmer is the average absolute difference between the amplitudes of consecutive periods, divided by the average amplitude~\cite{ref_shimmer}. These two types of features are interesting because they can capture instant-to-instant changes in frequency and amplitude of the signal, and are known to capture some differences between biological systems and machine systems \cite{gao2021detection}.     % \item Voiced/Unvoiced Ratio, which is the quotient of the length of speech and non-speech segments of an audio file\footnote{We trim the leading an trailing silence from that audio files beforehand in order to remove invalid learning shortcuts~\cite{muller2021speech}.}

\textbf{Speaker gender.} This is a binary feature for male vs. female gender of the speaker in the audio file. While this feature may seem naive, TTS pre-training or warm-up is often based on single-speaker data such as the highly popular female speaker in the LJSpeech Dataset~\cite{ljspeech:online}. The use of single-speaker data for pre-training could potentially cause bias in speech synthesis. 

\textbf{Duration and loudness.} The duration feature that we used was simply the duration of the audio signal on a per-file basis. The loudness in dBFS (db relative to the maximum possible loudness). A square wave at maximum amplitude will be roughly 0 dBFS (maximum loudness)~\cite{ref_loudness}.

\textbf{Signal amplitude, power, and energy.} For amplitude, we used the highest amplitude of the signal, with and without the conversion into dBFS (which specifies the value relative to the highest possible amplitude). We also calculated the power and energy over the entire audio file. 

\textbf{Noise ratio.} The mean and standard deviation of a short-term Harmonic to Noise Ratio (HNR) analysis. HNR is known to be predictive of audio deepfakes \cite{sanchez2015toward}.

\subsection{Neural Embeddings}\label{ss:nn_feats}
Alternatively, we can learn the attacker signature.
We design a neural network $f_\theta$
which maps a waveform $W$ to an attacker signature $f_\theta(W)$.
Following the architecture presented in~\cite{jia2018transfer}, this network consists of a stack of three recurrent layers.
Each layer in the stack consists of an LSTM with 768 neurons, whose output is fed into a 256-dimensional dense layer (called projection layer).
The output of the network is an embedding in $f_\theta(W) \in \mathbb{R}^{256}$.
The model is trained via the Angular Prototypical loss~\cite{angleproto}, which is a form of cosine similarity loss.

\section{Experimental Results}
\subsection{Evaluation Data}\label{ss:data}
% We use two sets of data to evaluate our attacker signatures proposed in \cref{s:methodology}.
We use the ASVspoof 2019 dataset~\cite{asvspoof2019challenge} to evaluate our attacker signatures that were proposed in Section~\ref{s:methodology}. 
Specifically, we use the \emph{Logical Access} (LA) part of ASVspoof 2019, which we abbreviate in this paper as: ASV19.
It consists of speech audio files which are either bonafide (i.e. authentic recordings of human speech) or spoofed audio (i.e. synthesized or fake audio).
The fake audio originates from 19 different attackers, labeled $A1$ - $A19$.
For each attacker, there are 4914 audio recordings, while there are 7355 bonafide samples.
ASV19 was the official dataset for the ASVspoof 2019 challenge
\cite{asvspoof2019challenge}, which focused on spoof detection, i.e. binary classification of authenticity.
A significant number of related work have used ASV19 data to develop detection algorithms~\cite{chintha2020recurrent, alzantot2019deep, chettri2019ensemble, wang2020densely}.

% Additionally, there is its successor~\cite{asvspoof2021challenge}, the ASVspoof 2021 dataset (again the LA part, abbreviated ASV21). 
% It is similar to the ASV19 dataset, but introduces more samples and new attackers.
% Additionally, it employs a variety of audio codecs, which blur the spoof's characteristics, making the challenge more difficult.
% For ASV21, the authors chose to not release the labels, which is why we only have access to the raw audio files, and not to their labels (i.e. neither bona-fide/spoof nor the individual attacker IDs). 
% This poses some challenges for our evaluation, c.f. \cref{sss:eval_21}.

\begin{figure}[t]
\centering
    \includegraphics[width=0.45\textwidth, trim={0 0 1.10cm 0},clip]{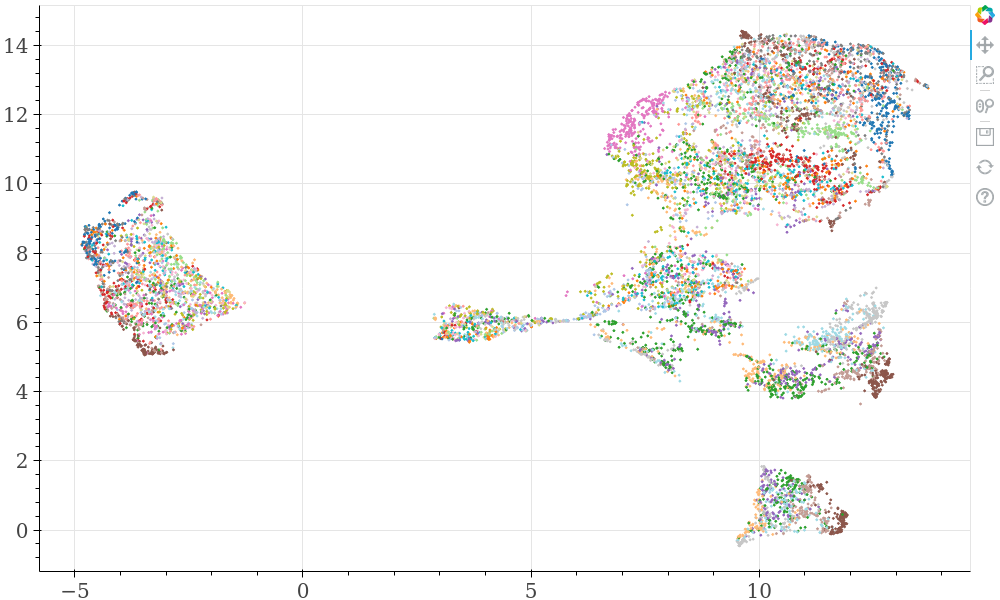}
    \caption{Clusters formed by our 16-dimensional embeddings from low-level signal features, using a 2D UMAP projection of the embedding space.
    The color of the data points indicates the different attacker ID labels.
    }
    \label{fig:asv19_sig}
\end{figure}

\subsection{Evaluation Metric}
\label{ss:ccvar}
To evaluate the quality of our clustering given a labeled dataset, we compute the average class-conditional variance\footnote{we estimate both mean and variance from a sample distribution and thus use Bessel-correction $N_c - 1$ in the denominator.} as follows in Equation~\ref{eq1} and Equation~\ref{eq2}:
% \[
%   mdcc(X, y=c) = \frac{1}{N_c} \sum_{i}^{N_c} | \mu_c - x^c_i | 
% \]
\begin{equation}
    var_C(X) = \frac{1}{|C|}\sum_{c \in C} var(X, y=c)
    \label{eq1}
\vspace*{-2mm}
\end{equation}
where
\begin{equation}
   var(X, y=c) = \frac{1}{N_c - 1} \sum_{i=1}^{N_c} \left( \overline{x}_c - x_c^{(i)} \right)^2
   \label{eq2}
\end{equation}

% \todo{var or bessel corrected var}
\noindent describes the class-conditional variance and 
$
\overline{x}_c = \sum_{i=1}^{N_c} x_c^{(i)}
$ is the centroid of class $c$; with $N_c$ instances $x_c^{(i)}$ belonging to class $c$.
Intuitively, this describes how closely packed the clusters are: the smaller $var_C$, the better the clustering w.r.t. to the target classes $C$. Note that this metric is sensitive to the scale of the data, which is why we consistently apply standard-normalization first. Lower values for class-conditional variance are indicative of better clustering.

\subsection{Evaluation of Low-Level Signal Features}\label{ss:eval_conventional}

% \begin{figure}[t]
%     \centering
%     \subfloat[\centering The \emph{jitter} feature, shown per attack ID.]{
%         \includegraphics[width=0.45\textwidth, trim={0 0 1.10cm 0},clip]{imgs/box_plots/box_plot_jitter}
%     }
%     \newline
%     \subfloat[\centering The \emph{pitch mean} feature, shown per attack ID. ]{
%         \includegraphics[width=0.45\textwidth, trim={0 0 1.10cm 0},clip]{imgs/box_plots/box_plot_pitch_mean}
%     }
%     \caption{The values of two static features, \emph{jitter} and \emph{pitch mean}, shown as box plot per attack ID.
%     The \emph{jitter} feature is the best feature in our conventional set, c.f.~\cref{tab:asv19_sig}, and is somewhat indiciative of the attack ID.
%     \emph{pitch mean} represents the class of features with a larger \emph{class-conditional} variance. 
%     It can be seen that this feature is about the same for all attack IDs.
%     }
%     \label{fig:box_plots}
% \end{figure}

%First, we evaluate the effectiveness of the conventional features (c.f. Section~\ref{ss:conventional_feats}) for attacker signature creation.
%For the ASV19 dataset, we compute the traditional signature candidates and use these as (standard-normalized) features to cluster the samples.
%We then evaluate the quality of the clusters, using the labels of ASV19.

Figure~\ref{fig:asv19_sig} shows how our 16-dimensional embeddings form clusters when projected into a 2D space using UMAP. The attacker ID labels are indicated by the colors of the data points. While some of the attack types are grouped together, there is no clear inter-label separation, and the clusters comprise many different attack IDs.
While the data seems to cluster locally, there is no strong separation between the clusters.
Rather, single clusters comprise multiple labels.
Thus the conventional features are limited in their applicability to attacker signature identification.
% \todo{is this understandable?}

\begin{figure}
    \includegraphics[width=0.45\textwidth, trim={0 0 1.10cm 0},clip]{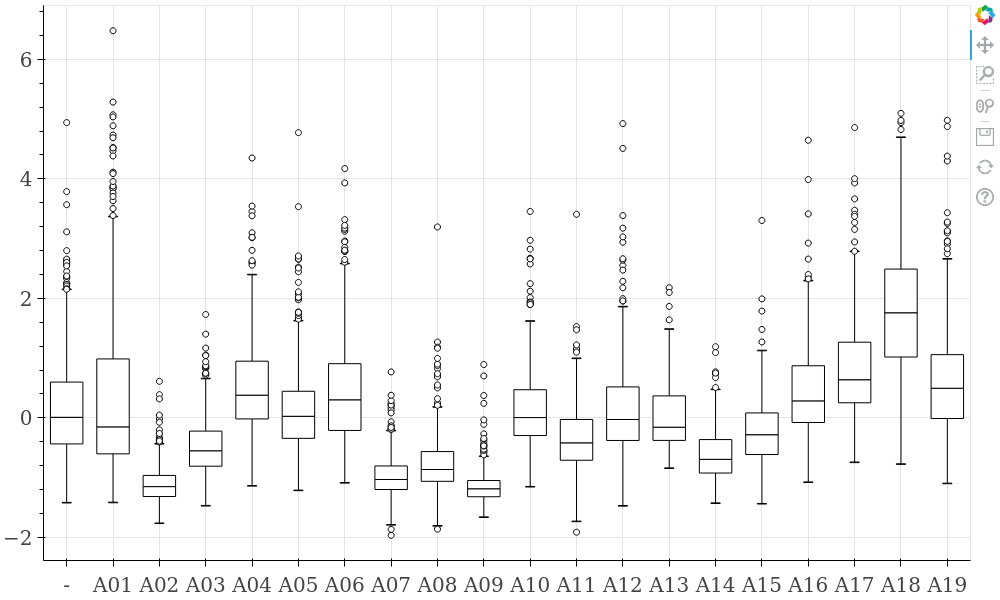}
    \caption{The values of the \emph{jitter} static features, shown as box plot per attack ID.
    While it is somewhat indicative of the class, the domain overlaps for many classes, which limits its suitability for attacker signature creation.
    }
    \label{fig:box_plots}
\end{figure}

\begin{table}[t]
    \vspace{2mm}
    \centerline{
        \begin{tabular}{lrr}
\toprule
{Low-Level} &  Class-Conditional \\
{Signal Feature} &  Variance (Avg.)\\
\midrule
duration       &           0.97 \\
energy         &           0.93 \\
gender          &           0.69 \\
HNR mean        &           0.61 \\
HNR std         &           0.71 \\
\cellcolor{LightCyan}\textbf{\emph{jitter}}          & \cellcolor{LightCyan}\textbf{\emph{0.53}} \\
loudness        &           0.67 \\
max loudness    &           1.21 \\
peak amplitude  &           1.21 \\
pitch MAS       &           0.72 \\
pitch max       &           0.85 \\
pitch mean      &           0.98 \\
pitch min       &           0.87 \\
pitch std       &           0.80 \\
power           &           0.73 \\
shimmer         &           0.73 \\ \midrule
all &           0.83 \\
\bottomrule
\end{tabular}

    }
    \caption{Evaluating the resulting clusters from our 16-dimensional embeddings created from low-level signal features. The average class-conditional variance is reported as averaged over all attack IDs. Lower values would indicate that a feature contributes to better clustering of attack IDs. 
    }
    \label{tab:asv19_sig}
\vspace*{-8mm}
\end{table}

We can verify this numerically by computing the average distance of each instance to its class centroid, as shown in Table~\ref{tab:asv19_sig}.
For reference, we have also included a box-plot of our strongest low-level signal feature, \textbf{\emph{jitter}}, which achieves a conditional-class variance of $0.53$. The jitter feature is somewhat indicative of attacker ID, as shown in Figure~\ref{fig:box_plots}.
To compute the values reported in Table~\ref{tab:asv19_sig}, we first compute the class-conditional variance and then compute the average over all attack IDs. 
All features are standard-normalized.
We see that the class-conditional variance is high. The usability of these features for attack signature clustering is very limited and the class-conditional values explain why clustering was not achieved in Figure~\ref{fig:asv19_sig}. The corresponding boxplot in Figure~\ref{fig:box_plots} shows that while \textbf{\emph{jitter}} is somewhat indicative of attacker ID, the ranges of values for jitter overlap significantly. Of all the low-level features, jitter was the best one based on our analysis. We conclude that our selected low-level features are of limited usefulness. Note that the majority of the features in Table~\ref{tab:asv19_sig} are even less useful than jitter. Overall, the class conditional variance is $0.83$ on average, which is much worse than jitter at $0.53$.

%\begin{figure*}[!ht]
%    \centering
%    \subfloat[\centering In-domain evaluation on ASV19.]{
%    \includegraphics[width=0.45\textwidth, trim={0 0 1.10cm 0},clip]{imgs/asv19_random_10_split.png}
%    % \caption{In-domain evaluation: A two-dimensional UMAP projection of the 256-dimensional embedding space obtained from our neural embedding model. The model has been trained on a random 10\% split of ASV19, and evaluated on the remaining 90\%, i.e. a disjoint test set (shown here). We see distinct clusters which clearly correspond to the attacker-ID, as indicated by the color.}
%    \label{fig:in_domain19}
%    }
%    % \newline
%    \subfloat[\centering Out-of-domain evaluation on ASV19.]{
%    \includegraphics[width=0.45\textwidth, trim={0 0 1.10cm 0},clip]{imgs/asv19_4_speaker_split.png}
%    \label{fig:out_of_domain19}
%    }
%    \caption{Evaluation of our neural embedding attack signatures on ASV19. \emph{Left (a):} In-domain task where all attacker types were seen during training. \emph{Right (b):} Out-of-domain task where four randomly chosen attacker types were completely held-out from training. }
%    
%    \label{fig:out_of_domain19}
%\end{figure*}

%\subsubsection{Downstream Evaluation}\label{sss:neural_eval_conv}
Another method that we use to evaluate the features is through supervised classification. We train a classifier to predict the attacker ID, given the 16-dimensional feature vector.
We choose a simple feed-forward network with 3 hidden layers, $50$ neurons per layer, ReLU activation~\cite{agarap2018deep}, and learning rate of $0.001$. We train on $90\%$ of the ASV19 data while evaluating on the other $10\%$ as held-out test data.
We achieve an accuracy of \accstatic/\%, which is significantly better than random guessing (the random baseline has $5\%$ accuracy, given $20$ different attacker IDs), but inadequate still. We conclude that the low-level features from Section~\ref{ss:conventional_feats} are not suitable for attacker signatures.

\subsection{Evaluation of Neural Embeddings}\label{ss:eval_nn}
We present our evaluation of neural embeddings for attacker signature (c.f. Section~\ref{ss:nn_feats}) evaluated on the ASV19 dataset.
% \subsubsection{Evaluation on labeled data}\label{sss:eval_19}
% First, we evaluate the ASV19 data, which is fully labeled.
We evaluate two different scenarios: \emph{in-domain} and \emph{out-of-domain}. We describe each scenario and present clustering results for each one. We also describe how well the attack signature embeddings perform in a separate classification task.
%\subsection{In-Domain and Out-of-Domain Scenarios}
For the \emph{in-domain} scenario, we train a neural network to learn embeddings using the training partition of the ASV19 dataset. Our network is trained on 90\% of this partition and evaluated on the remaining 10\%. By treating this 10\% as our test data, we are able to evaluate our embeddings using new audio recordings from attackers seen during training. 
For the \emph{out-of-domain} scenario, we select four randomly chosen attacker IDs (A02, A04 A012, A14) and reserve all of their audio samples as a test set. We then train on the remainder of the data. This allows us to evaluate our ability to attribute unseen audio recordings from unknown attackers, as we would expect in a real-world application such as a tool for cybersecurity purposes.

\begin{figure}[t]
    \centering
    \includegraphics[width=0.45\textwidth, trim={0 0 1.10cm 0},clip]{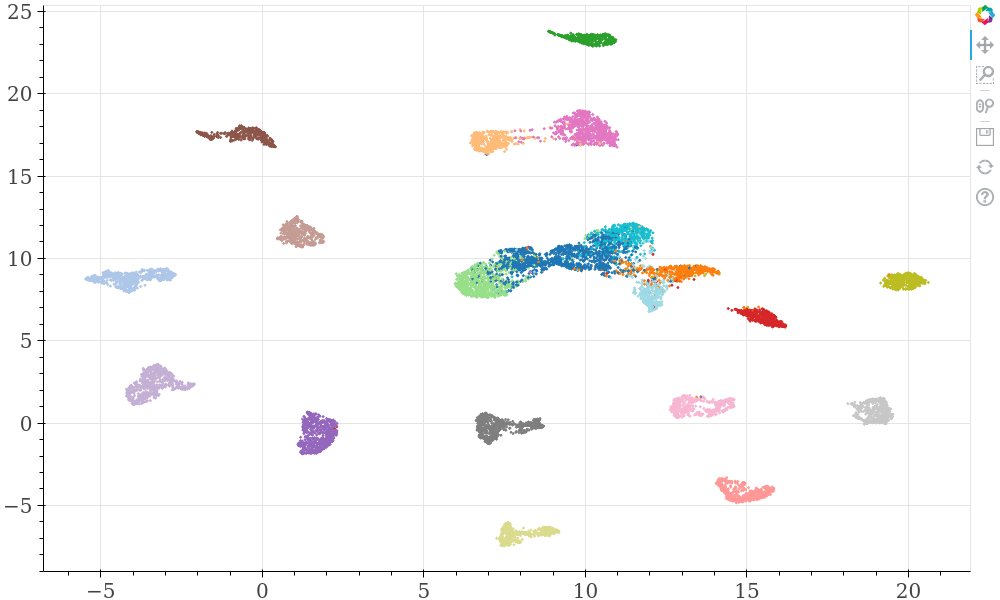}
    % \caption{In-domain evaluation: A two-dimensional UMAP projection of the 256-dimensional embedding space obtained from our neural embedding model. The model has been trained on a random 10\% split of ASV19, and evaluated on the remaining 90\%, i.e. a disjoint test set (shown here). We see distinct clusters which clearly correspond to the attacker-ID, as indicated by the color.}
    \caption{In-domain evaluation of our neural attack signatures on ASV19. All attacker types were seen during training.}
    \label{fig:in_domain19} 
\end{figure}

\begin{figure}[t]
    % \newline
    \includegraphics[width=0.45\textwidth, trim={0 0 1.10cm 0},clip]{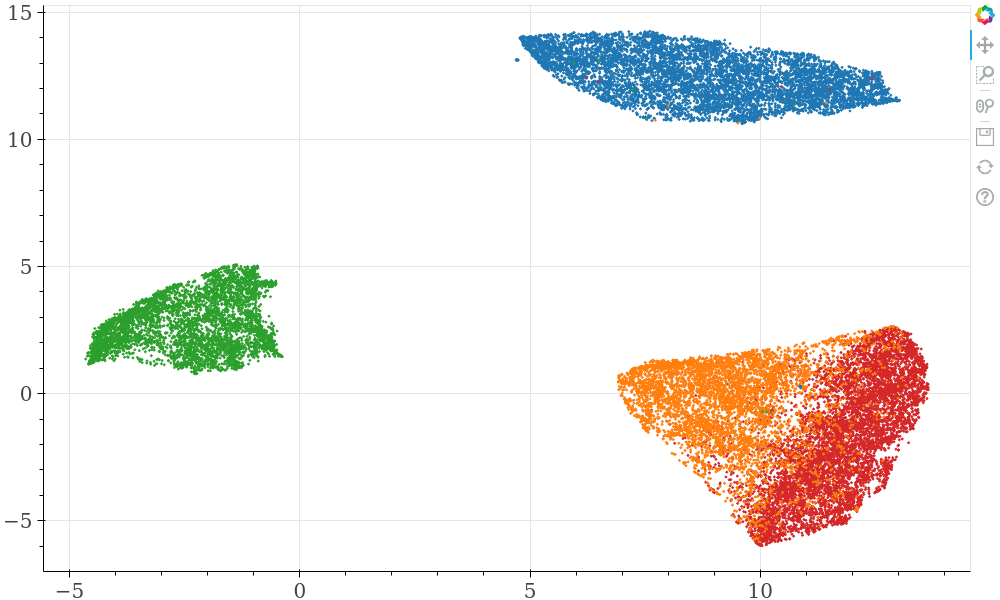}
    \caption{Our-of-domain evaluation of our neural attack signatures on ASV19. Four attackers were hld-out from training. }
    \label{fig:out_of_domain19}
\vspace*{-4mm}
\end{figure}

The resulting clusterings for each scenario are shown in Figure~\ref{fig:in_domain19} and Figure~\ref{fig:out_of_domain19}.
As expected, we observe that the \emph{in-domain} task seems easier than the \emph{out-of-domain} task due to the successful separation of the large number of clusters. Though in both cases we obtain sensible clustering. Computation of the average class-conditional variance yields $0.19$ for the \emph{in-domain} task and $0.46$ for the \emph{out-of-domain} task. Both of these significantly outperform the clustering evaluation from low-level signal features, which had an average class-conditional variance of $0.83$. The out-of-domain scenario is more challenging, since the audio file content and attackers are unseen during training. Nevertheless, we obtain very clear clustering, indicating the our neural embedding signatures can generalize beyond known attacker IDs and may be usable in a real-world scenario.

%\subsection{Downstream Evaluation with ASV21 Data}
Additionally, we evaluate the neural embeddings as a downstream classification task. We again create a simple a three-layer feed-forward neural network with ReLU activation~\cite{agarap2018deep}, using 50 neurons per layer, and a learning rate of $0.001$. The model takes as input our previously computed neural embeddings as the attack signatures. We use our signatures as feature vectors while attempting to classify attack types. We train on $90\%$ of the ASV19 data while evaluating on the other $10\%$. Our simple classifier achieves a test accuracy of \accneural/\% for the \emph{in-domain} evaluation.
%\footnote{Note that we cannot apply classification to the \emph{out-of-domain} scenario since the classification targets need to be seen during training using this method.}. 
The classification accuracy achieved by our neural embedding attack signatures (\accneural/\%) significantly outperforms what we had achieved with our low-level signal feature attack signatures (\accstatic/\%). 

% \subsubsection{Summary}
% In this section, we evaluate our neural embeddings on both labeled and unlabeled data.
% We show that the clustering is strong when evaluated on labeled data, even for the difficult \emph{Out-of-Domain} task.
% When evaluated on unlabeled data, we find high visual correspondence of the clusters to known attacker IDs; and we find that the clusters are not meaningful to irrelevant features such as gender of pitch.

%\section{Future Work}\label{sss:eval_21}
\begin{figure}[t]
    \centering
    \includegraphics[width=0.45\textwidth, trim={0 0 1.10cm 0},clip]{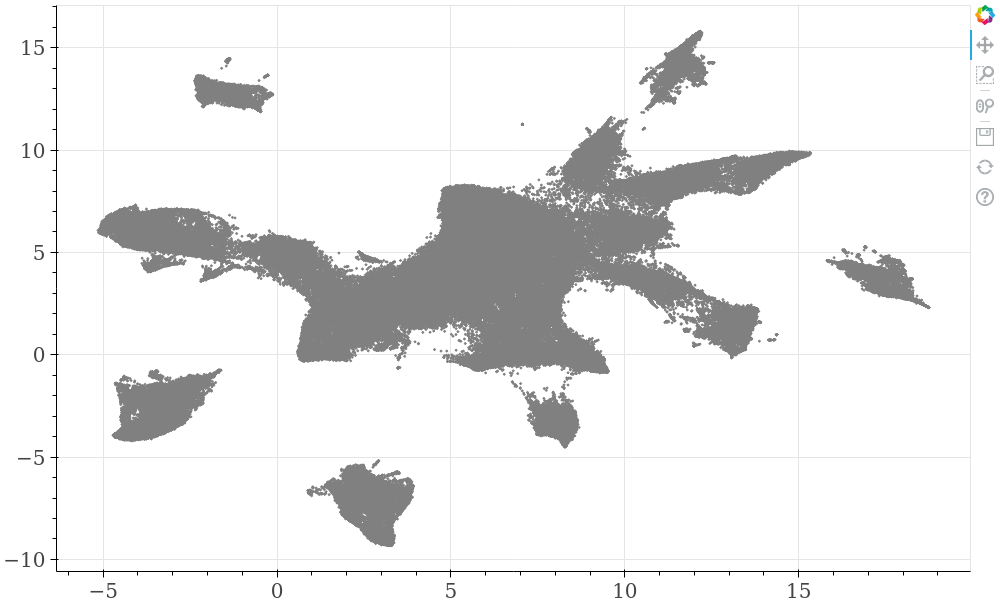}
    \caption{Visualisation of the unlabeled ASV21 data via 2D UMAP projection to cluster our neural attack signatures.}
    \label{fig:asv21}
\vspace*{-4mm}
\end{figure}

We applied our neural embedding attack signature technique to a completely held-out dataset, from the ASVspoof 2021 challenge (ASV21). It is similar to the ASV19 dataset but introduces more samples, new attackers, and a variety of codecs which blur the characteristics of spoofed audio that our model trained on. The attacker labels have not yet been released therefore we leave to future work an extensive evaluation once the labels have been released. we can apply our ASV19-trained model to the dataset and inspect the unlabeled clustering.
This is shown in \cref{fig:asv21}.
We can see that our model succeeds in finding various clusters, which indicates that our model generalizes beyond the ASV19 dataset.

\section{Conclusion and Future Work}
In this paper we introduced two new methods for creating attacker signatures to attribute spoofed audio to specific attackers. We evaluated signatures from low-level signal features and neural embeddings. We found that neural embeddings are well-suited for this problem and perform better in clustering and classification than low-level signal features. 
%statically-computed candidates such as the audio's pitch, loudness, etc., and find that these do not suffice to generate attacker signatures.
%We then present a neural embedding which we evaluate on two large corpora of spoofed audio.
%Using the available ground-truth labels, we show that the presented neural attacker signature can be used to identify and group together fakes from both seen and unseen attackers.
We have also demonstrated how our methodology can be applied to a completely held-out dataset and we suggest future work to investigate model adaptation once the labels have been released for the ASV21 dataset. 
%and apply our model to a large, unlabeled dataset, the ASV21 dataset.
Our model produces distinct clusters, indicating that it has learned to generalize beyond the ASV19 dataset. The neural embedding attack signatures are a very promising avenue for future attacker attribution research.
% and that the semantics of these clusters are meaningful.

\bibliographystyle{IEEEtran}
    
\bibliography{mybib}

\end{document}